# Automating API Documentation with LLMs: A BERTopic Approach

AmirHossein Naghshzan
amirhossein.naghshzan.1@ens.etsmtl.ca
École de Technologie Supérieure
Montreal, Quebec, Canada

**Abstract**

Developers rely on API documentation, but official sources are often lengthy, complex, or incomplete. Many turn to community-driven forums like Stack Overflow for practical insights. We propose automating the summarization of informal sources, focusing on Android APIs. Using BERTopic, we extracted prevalent topics from 3.6 million Stack Overflow posts and applied extractive summarization techniques to generate concise summaries, including code snippets. A user study with 30 Android developers assessed the summaries for coherence, relevance, informativeness, and satisfaction, showing improved productivity. Integrating formal API knowledge with community-generated content enhances documentation, making API resources more accessible and actionable work.

## 1 Introduction

Developers frequently use APIs but often struggle with lengthy, complex, or incomplete documentation [1]. Many turn to Stack Overflow and GitHub for quick solutions [2]. However, these sources are unstructured and vast. Automating the extraction and summarization of key insights can enhance documentation usability.

Previous work has used Stack Overflow data to help developers [4, 9]. Our work goes further by addressing a different challenge: providing concise summaries of problem-solution discussions, and offering a broader understanding beyond specific code solutions.

Our approach leverages NLP techniques to extract and summarize frequent API-related issues from Stack Overflow, focusing on Android APIs. BERTopic [3], a topic modeling technique that combines BERT embeddings [5] and class-based TF-IDF (c-TF-IDF), to identify prevalent topics discussed in Android-related Stack Overflow posts. We then use extractive summarization methods to generate concise summaries of frequent problems and potential solutions within these topics.

To guide our research, we seek to answer the following research questions:

**RQ1**: *What key challenges do developers face with Android APIs based on Stack Overflow discussions?*

**RQ2**: *Can automated summarization effectively capture and summarize these challenges?*

Our goal is to provide developers with an efficient means to access summarized knowledge derived from community discussions. This research lays the groundwork for practical tools, such as IDE plugins, to improve developer productivity.

## 2 Methodology

Our research methodology consists of three main steps: data collection and pre-processing, topic modeling with BERTopic, and summarization with BERT.

### 2.1 Data Collection and Pre-processing

We utilized the Stack Exchange API to retrieve all the questions on Stack Overflow that were tagged with *Android* from January 2009 to April 2023. We chose Android because it ranks among the top eight most frequently discussed subjects on Stack Overflow [6]. To prepare our data for further analysis, we followed these pre-processing steps: lowercasing the text, removing stop words and punctuation, tokenizing the text, lemmatizing the tokens, stemming the tokens, and finally removing special characters and numerical values.

### 2.2 Topic Modeling with BERTopic

To identify the topics discussed in Stack Overflow, we utilized the BERTopic (Bidirectional Encoder Representations from Transformers Topic Modeling) [3]. BERTopic, leveraging BERT embeddings, clusters semantically related posts. We implemented BERTopic using a pre-trained model from Hugging Face[1], running on Google Colab Pro, which comes equipped with a T4 GPU, boasting 25GB of VRAM and 26GB of system RAM. UMAP [8] reduced embedding dimensions for visualization, and HDBSCAN [10] clustered topics. Figure 1 illustrates the word scores for the top 12 topics. These scores indicate the relevance of specific words to each topic, with higher scores representing words that are more characteristic of their respective topics.

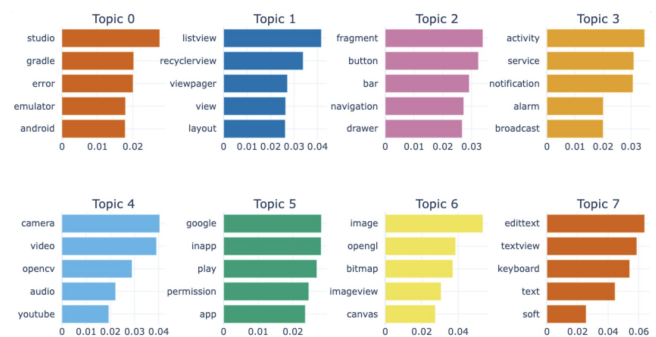

**Figure 1: Word scores of Stack Overflow's Android topics**

---

[1] https://huggingface.co

Table 1: Generated summaries for questions and answers of topic project_error_build_gradle

| Problem | Solution |
|---|---|
| Jenkins tries to launch tools instead of emulator I'm trying to set up jenkins ui tests and it fails on running emulator command | I'm answering my own question its an issue with android emulator plugin 3.0 not working with "Command line tools only" sdk package. |
| I am trying to add Kotlin sources of an AAR in Android Studio 3.3.2. It doesn't work when I select "Choose Sources" and choose the corresponding source.jar but it only displays "Attaching" | Configure kotlin in your project following these steps in build.gradle. Also add classpath org.jetbrains.kotlin gradle plugin clean and build your project it worked for me. |

## 2.3 Summarization with BERT

Previously, Naghshzan *et al.* [6, 7] demonstrated that extractive summarization can be employed to extract information from Stack Overflow. Therefore, we utilize the extractive version of BERT [11]. We are interested in identifying frequent problems that developers face when dealing with APIs and recommending potential solutions. Therefore, our summarization approach involved:

(1) Summarizing frequent issues via high-scoring questions (3 votes) within each topic.
(2) Extracting accepted or highly rated answers (2 votes) for concise solution summaries.
(3) Including relevant code snippets for practical guidance.

As an example, Table 1 presents summaries for topic 0 (*project*), highlighting two frequent problems and their solutions, using this dual approach of topic modeling and summarization.

## 3 Survey & User Study

We conducted a user study with 30 Android developers, stratified by skill level: beginner (n=10), intermediate (n=12), and advanced (n=8). Participants evaluated the summaries based on four key criteria: **coherence** (logical and clear presentation), **informativeness** (usefulness of the content), **relevance** (alignment with the topic), and **overall satisfaction**. Ratings were provided on a 5-point Likert scale ranging from *Strongly Disagree* (1) to *Strongly Agree* (5).

The results demonstrated positive feedback across all criteria, with average scores of 3.2 out of 4 for coherence, 3.1 for informativeness, 3.3 for relevance, and 3.4 for overall satisfaction. These findings suggest that the summaries were well-received, effectively capturing the key issues and solutions, and offering practical value to participants.

## 4 Discussion

Our study demonstrates that summarizing informal documentation sources like Stack Overflow can significantly aid developers in understanding frequent API-related issues. By providing concise summaries, developers can quickly identify prevalent problems and solutions without sifting through numerous posts.

While Stack Overflow already offers extensive information, our approach complements it by aggregating and summarizing discussions across posts, providing an efficient starting point for developers encountering new issues. These summaries, integrated into tools like IDEs, can save time by highlighting frequent problems, actionable solutions, and relevant code snippets, enhancing productivity and accessibility.

### 4.1 Potential Risks and Limitations

Summarization inherently risks omitting nuances and conflicting viewpoints, as it condenses information. Furthermore, our approach's reliability depends on the variable quality of Stack Overflow content. The current focus on Android APIs also limits generalizability to other domains. Finally, the small sample size (n=30) necessitates larger-scale validation to confirm these findings.

### 4.2 Future Research

Future work includes extending our approach to other programming languages and APIs, exploring hybrid summarization for richer output, incorporating user feedback for personalization, and conducting large-scale user studies to validate and generalize findings.

## 5 Conclusion

Developers struggle with complex API documentation and informal sources like Stack Overflow. This research presents a novel approach using BERTopic and BERT-based summarization to automatically extract and condense common Android API problems and solutions from Stack Overflow. Our approach identifies prevalent topics and provides developers with concise, community-derived information. A user study (n=30) showed the summaries are generally coherent, informative, and helpful. This work demonstrates the potential of NLP for processing informal documentation. Future work will refine summarization, address limitations like conflicting snippets, and integrate summaries into developer tools.